\documentclass[nojss]{jss}

\usepackage{thumbpdf,lmodern}

\usepackage{graphicx}

\newcommand{\class}[1]{`\code{#1}'}
\newcommand{\fct}[1]{\code{#1()}}

\author{Mark P.J. van der Loo\\Statistics Netherlands}
\Plainauthor{Mark P.J. van der Loo}

\title{Monitoring data in \proglang{R} with the \pkg{lumberjack} package}
\Plaintitle{Monitoring data in R with the lumberjack package}
\Shorttitle{Monitoring data in \proglang{R} with the \pkg{lumberjack} package}

\Abstract{
Monitoring data while it is processed and transformed can yield detailed
insight into the dynamics of a (running) production system.  The
\pkg{lumberjack} package is a lightweight package allowing users to follow how
an \proglang{R} object is transformed as it is manipulated by \proglang{R}
code. The package abstracts all logging code from the user, who only needs to
specify which objects are logged and what information should be logged. A few
default loggers are included with the package but the package is extensible
through user-defined logger objects. 
}

\Keywords{Data Quality, Process Monitoring, Logging, Debugging, \proglang{R}}
\Plainkeywords{Data Quality, Process Monitoring, Logging, Debugging, R}

\Address{
  Mark P.J. van der Loo\\
  \href{https://orcid.org/0000-0002-9807-4686}{\includegraphics[keepaspectratio=true, height=\fontcharht\font`\0]{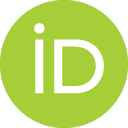}} \url{https://orcid.org/0000-0002-9807-4686}\\
  Research and Development\\
  Statistics Netherlands\\
  Henri Faasdreef 312\\
  2492JP Den Haag, The Netherlands\\
  E-mail: \email{m.vanderloo@cbs.nl}
}

\usepackage{Sweave}
\begin{document}

\section{Introduction}
It is common practice to monitor a data analyses process while it is running.
Especially in production environments where analyses are run repeatedly on
different but structurally comparable data sets. Following a running procedure
is usually done with some form of logging system, where the running process
updates a log that can be tracked by users as it proceeds.

One can distinguish two types of monitoring. On the one hand there is
\emph{process logging}, or just \emph{logging} for short. Here, the running
system notifies users of progress and significant events, usually by writing
short time-stamped messages to a file (where `file' can be a flat text file,
database, screen or any other device accepting such input). The aim of these
messages is to signal whether procedures have concluded successfully, and if
they haven't, to report what went wrong. Such information is highly valuable in
\emph{post-mortem} investigations, for example when a production script has
crashed. On the other hand there is \emph{tracing} where the state of variables
is followed over the course of the process. Tracing is usually applied at the
development stage as a debugging tool, often using an interactive interface
tool to run the code line by line while inspecting the state of variables. One
of the purposes of this paper is to demonstrate that targeted forms of
automated tracing can be useful at the production stage as well.

The ability to trace the state of variables for debugging purposes is common
across languages for technical or statistical computing. Focussing on
\proglang{julia} \citep{bezanson2017julia}, \proglang{python}
\citep{rossum2009python}, and \proglang{R} \citep{rcore2019}, we see that all
have this capability built into their standard libraries. In \proglang{julia},
the \pkg{Debugger} module provides ways to set break points that allow
programmers to investigate the scope of a running function at that point, to
browse the call stack, and to execute the code step-by-step. Similar
functionality is offered in \proglang{python} through the \pkg{pdb} module and
by \proglang{R}'s \pkg{base} and \pkg{utils} packages. Although there are some
differences, the functionality across these languages is comparable. 

When it comes to process logging, \proglang{R} differs significantly from
\proglang{julia}  or \proglang{python}. The latter two languages offer a
logging module as part of their standard library, respectively called
\pkg{Logging} and \pkg{logging}. In \proglang{julia}, the \pkg{Logging} module
offers a mechanism that is somewhat comparable to how exceptions are handled:
programmers can insert logging statements throughout their code and use default
or self-written local or global handlers to process and store log messages.
Logging handlers are organized in a type hierarchy where the `root' handler
ultimately handles all logging messages that are not taken care of by
lower-level loggers. This is similar to how logging is organized in
\proglang{python}'s \pkg{logging} module. One difference is that in
\proglang{python} the logging configuration, including logging level and output
file (via \code{logging.basicConfig()}) can be set only once per session.

\proglang{R} has no native logging mechanism, but for process logging several
packages are available via CRAN\footnote{\url{https://cran.r-project.org}}.
The two most popular ones by far\footnote{Based on download statistics obtained
with \pkg{dlstats} by \citet{yu2019dlstats}.} are currently \pkg{futile.logger}
\citep{rowe2016futile}, and \pkg{logging} \citep{frasca2019logging}.  Other
implementations include \pkg{logger} \citep{daroczi2019logger}, \pkg{loggit}
\citep{price2018loggit}, \pkg{log4r} \citep{myles2020log4r}, and \pkg{rsyslog}
\citep{jacobs2018rsyslog}. Typical features for these packages include the
ability to distinguish between different classes of messages, setting a logging
level (threshold) that decides which messages are created at runtime, and
customizing output messages.  Typical message types include `information',
`warning', `error', and sometimes `debug'.  When comparing the functionality of
these packages, \pkg{futile.logger}, \pkg{logging}, and \pkg{logger} are
especially similar as all of them are inspired by a \proglang{Java} logging
system called
\pkg{log4j}\footnote{\url{https://logging.apache.org/log4j/2.x/}}. This system
is again similar to \proglang{julia}'s \pkg{Logger} or \proglang{python}'s
\pkg{logging}, with a configurable hierarchy of log handlers. The three
\proglang{R} packages mainly differ on details such as the granularity of
available logging thresholds (\pkg{logging} has the most), available output
channels (\pkg{logger} offers the most), and look-and-feel. The \pkg{loggit}
package distinguishes itself by offering dedicated logging of non-standard
conditions: it sends error, warning, and message conditions to \proglang{JSON}
output as well as passing them through to \code{stderr}. Finally, the
\pkg{rsyslog} package is written to resemble an operating system's
\code{syslog} interface.  On POSIX complient operating systems, all logging
messages are send to the central \code{syslog} file.

Between the possibilities of interactive variable tracing and process logging
during production runs there seems to be a gap in functionality where the state
of variables is traced automatically while running in production. Such
functionality may serve interesting use cases. For example, consider a
frequently running production system that includes elaborate data cleaning,
imputation, and transformation steps. It is interesting to monitor the effect
that each step has on the variables, both to understand their relative
importance within the whole procedure, and to monitor changes in this relative
importance over production runs. Significant changes over time may indicate
that data circumstances have changed to the extent that assumptions upon which
the data processing is developed may need to be reconsidered. At the
development stage, such monitoring can help deciding whether the contribution
of each processing step is worth the extra complexity and runtime of the whole
procedure.

The \pkg{lumberjack} package \citep{loo2020lumberjack} presented in this paper
aims to fill this gap between interactive tracing and process logging. It
allows users to specify which objects should be monitored and how. Users can
either follow a (summary of) the state of an object or measure differences
between consecutive versions of an object as it gets processed. For example,
one can follow the average of a variable in a data frame that gets processed or
count the number of cells that changed after each operation. In the simplest
case this can be done by adding just a single line code to an existing
\proglang{R} script.

The package is designed with three core design principles in mind.  First, a
user should not have to worry about data monitoring while developing the main
process. Ideally, a user develops a production script and later simply adds a
specification stating which variables to monitor and how.  This means that the
package should \emph{separate concerns} between developing a production script
and monitoring data. Second, the monitoring process should neither require any
change in user code, nor rely on behaviour of code used from  other packages:
monitoring must be \emph{agnostic} with respect to the code that actually
processes the data. Third, the package must allow users and developers complete
\emph{flexibility} in how to track changes in data. Depending on the objects
that are followed, many different parameters may be interesting and the package
must therefore be extensible with user-defined monitoring capabilities. 

The following Section demonstrates how to monitor \proglang{R} objects with
\pkg{lumberjack}, both in batch and in interactive mode. In
Section~\ref{sect:custom} it is shown how the package can be extended with
custom loggers by users or package developers. A conclusion is given
in Section~\ref{sect:conclusion}.

\section{Monitoring R objects}
\label{sect:monitoring}
In what follows a running example will be used based on the `supermarkets'
data set that is included in the supplementary materials.  The data set is
derived from the \code{retailers} data set of the \pkg{validate} package
\citep{loo2019validate}.
\begin{Schunk}
\begin{Sinput}
R> head( read.csv("supermarkets.csv"), 3 )
\end{Sinput}
\begin{Soutput}
     id staff turnover other.rev total.rev
1 SPM01    75       NA        NA      1130
2 SPM02     9     1607        NA      1607
3 SPM03    NA     6886       -33      6919
\end{Soutput}
\end{Schunk}
Besides an identifying variable in the first column it contains `staff'
numbers, `turnover', `other revenue' and `total revenue' in kEUR of
sixty establishments.

\subsection{Monitoring changes in production scripts}
\label{sect:scripts}
A script called \code{supermarkets.R} shown in Figure~\ref{fig:supermarkets}
will serve as example production script. It reads \code{supermarkets.csv} and
then imputes and corrects `other revenue' values where deemed necessary. Next
it uses a ratio estimator to impute `staff' numbers based on `turnover'
amounts.  Finally, it derives a new variable called `ratio' holding the ratios
between `turnover' and `total revenue' and then writes the output to a new CSV
file.  In production circumstances such a file could be run using
\code{source("supermarkets.R")} or as follows while invoking \proglang{R}.

\begin{Code}
  R -q -f supermarkets.R
\end{Code}

\begin{figure}[!t]
\hrule
\begin{Sinput}

  spm <- read.csv("supermarkets.csv")

  # assume empty values should be filled with 0
  spm <- transform(spm
          , other.rev = ifelse(is.na(other.rev), 0, other.rev))

  # assume that negative amounts have only a sign error
  spm <- transform(spm, other.rev = abs(other.rev))

  # ratio estimator for staff conditional on turnover
  Rhat <- with(spm
          , mean(staff, na.rm = TRUE)/mean(turnover, na.rm = TRUE))

  # impute 'staff' variable where possible using ratio estimator
  spm <- transform(spm
        , staff = ifelse(is.na(staff), Rhat*turnover, staff))

  # add a column
  spm <- transform(spm, ratio = turnover/total.rev)

  # write output
  write.csv(spm, "supermarkets_treated.csv", row.names = FALSE)
\end{Sinput}
\hrule
\caption{A script that reads, transforms and writes the \code{supermarkets}
dataset (\code{supermarkets.R} in the supplementary materials).}
\label{fig:supermarkets}
\end{figure}

To track all possible changes in the supermarket data, a user assigns one or
more \emph{loggers} to existing \proglang{R} objects.  Here this is done by
adding a single line at the beginning of the script, just after reading the
\code{supermarkets.csv} file. The function \fct{start\_log} accepts a variable
name and a logging object which will be discussed below in more detail.
\begin{Sinput}
  spm <- read.csv("supermarkets.csv")

  start_log(spm, cellwise$new(key = "id"))

  # the rest of the script as in Figure 1.
\end{Sinput}
The altered script is stored as \code{supermarkets\_logged\_1.R} in the
supplementary materials.  Now, from a running \proglang{R} session (interactive
or in batch mode) the script must be executed as follows.
\begin{Schunk}
\begin{Sinput}
R> library(lumberjack)
R> out <- run_file("supermarkets_logged_1.R")
\end{Sinput}
\end{Schunk}
Alternatively one can run the script when invoking an \proglang{R} session as
follows.
\begin{Code}
  R -q -e 'library("lumberjack"); run_file("supermarkets_logged_1.R")'
\end{Code}

The function \fct{run\_file} has executed the script and signals that a log
file was written to \code{smp\_cellwise.csv} (the reason that \fct{run\_file}
is needed is discussed at the end of this Section). 
\begin{Schunk}
\begin{Sinput}
R> spm_log <- read.csv("spm_cellwise.csv") 
R> head(spm_log, 3)
\end{Sinput}
\begin{Soutput}
  step                     time                      srcref
1    2 2020-05-08 15:24:36 CEST supermarkets_logged_1.R#7-7
2    2 2020-05-08 15:24:36 CEST supermarkets_logged_1.R#7-7
3    2 2020-05-08 15:24:36 CEST supermarkets_logged_1.R#7-7
                                                               expression
1 spm <- transform(spm, other.rev = ifelse(is.na(other.rev),0,other.rev))
2 spm <- transform(spm, other.rev = ifelse(is.na(other.rev),0,other.rev))
3 spm <- transform(spm, other.rev = ifelse(is.na(other.rev),0,other.rev))
    key  variable old new
1 SPM01 other.rev  NA   0
2 SPM02 other.rev  NA   0
3 SPM06 other.rev  NA   0
\end{Soutput}
\end{Schunk}
Reading the log file yields a step count, a time stamp, a source reference, the
code that was executed, the key of the record where changes took place, the
name of the variable, and the old and the new value. As suggested by the name
of the logger (\code{cellwise}) it records changes cell by cell. For example,
in record \code{SPM06} the value of variable \code{other.rev} was altered from
\code{NA} to \code{0} by the \code{transform} expression shown in the third
column.

When a user just adds the single \fct{start\_log} expression, \pkg{lumberjack}
makes a number of default choices. These include the point where the logging
stops (after all the expressions in the \proglang{R} script have been executed)
and where the logging information is written. In the case of the
\code{cellwise} logger, both can be controlled by adding a line like
\begin{Sinput}
  stop_log(spm, file = "my_custom_log.csv")
\end{Sinput}
at the point where logging should stop. An overview of logging control
functions is given in Table~\ref{tab:loggers}. The fact that \fct{stop\_log}
accepts a \code{file} argument actually depends on the fact that \code{spm} is
tracked by the \code{cellwise} logger: not all loggers necessarily write
something to a file. The structure of loggers is discussed in more detail in
Section~\ref{sect:custom} but briefly, the loggers that come with
\code{lumberjack} are \code{R6} reference objects\footnote{Based on the
\code{R6} package of \citet{chang2019r6}}.  This means that the expression
\begin{Sinput}
  cellwise$new(key = "id")
\end{Sinput}
returns a new logger, that uses variable \code{id} as key variable. Not all
loggers need to know about a key and in fact the arguments given to \fct{\$new}
depend on the logger. An overview of loggers currently available in
\pkg{lumberjack} is given in Table~\ref{tab:loggers}.

\begin{table}[!t]
\centering
\caption{Logging control.} 
\label{tab:logcontrol}
\begin{tabular}{ll}
\hline
Logger          & what it does \\
\hline
\code{start\_log}   & Assign a logger to an \proglang{R} object.\\
\code{stop\_log}    & Stop logging and dump log, where dumping can be switched off.\\
\code{dump\_log}    & Dump logging info and stop logging, where stopping is optional\\
\code{run\_file}    & Execute a file, while logging, in a  new \code{environment}.\\
\code{source\_file} & Execute a file, while logging, in the global \code{environment}.\\
\code{\%L>\%}       & Pipe operator that also triggers logging where indicated.\\
\hline
\end{tabular}
\end{table}

\begin{table}[!t]
\centering
\caption{Loggers in \pkg{lumberjack}.}
\label{tab:loggers}
\begin{tabular}{ll}
\hline
Logger          & what it does \\
\hline
\code{expression\_logger} & record result of custom \code{R} expressions.\\
\code{filedump}           & dump a file after each operation.\\
\code{simple}             & record whether anything changed (\class{logical}).\\
\code{cellwise}           & record cell-by-cell changes.\\
\hline
\end{tabular}
\end{table}

To demonstrate the possibility of multiple tracking, two loggers tracking
the \code{spm} variable are specified so the top of the script
in Figure~\ref{fig:supermarkets} now looks like this.
\begin{Sinput}
  spm <- read.csv("supermarkets.csv")

  start_log(spm, logger = cellwise$new(key="id"))

  logger <- expression_logger$new(
                mean_staff     = mean(staff, na.rm = TRUE)
              , mean_other.rev = mean(other.rev, na.rm = TRUE) 
            )
  start_log(spm, logger=logger)
  # the rest of the script...
\end{Sinput}
The altered script is provided as \code{"supermarkets_logged_2.R"} in the
supplementary materials. Here, the mean of variables `staff' and `other.rev'
are tracked as the dataset is manipulated by the script. Running the file now
yields two messages, one for each logger.
\begin{Schunk}
\begin{Sinput}
R> run_file("supermarkets_logged_2.R")
\end{Sinput}
\end{Schunk}
Below the new log file is read, yielding a complete view on how the means of
`staff' and of `other revenue' vary as the data gets processed (the `srcref'
and `expression' columns are suppressed for brevity).
\begin{Schunk}
\begin{Sinput}
R> read.csv("spm_expression.csv")[c("step", "mean_staff", "mean_other.rev")]
\end{Sinput}
\begin{Soutput}
  step mean_staff mean_other.rev
1    1   11.53704      22.366792
2    2   11.53704       8.946717
3    3   11.53704      10.046717
4    4   11.53704      10.046717
5    5   12.07457      10.046717
6    6   12.07457      10.046717
7    7   12.07457      10.046717
\end{Soutput}
\end{Schunk}

Again, \pkg{lumberjack} chooses default places to stop logging and to dump the
logging data. The user can control this by inserting \fct{stop\_log} anywhere
in the code after logging started. It is possible to stop individual loggers 
with the \code{logger} argument. For example, to stop the cellwise logger at a
certain point, add the following.
\begin{Sinput}
  stop_log(spm, logger = "cellwise")
\end{Sinput}
This will dump the \class{cellwise} log for \code{spm} and stop using the
cellwise logger, but it will continue logging with the expression logger.  Note
that the combination of a variable name and a logger type is sufficient to
uniquely identify a logger instance: it is pointless to track the same object
with the same type of logger twice, and this is therefore not allowed by
\pkg{lumberjack}.

Summarizing, the interface implemented by the package consists of two main
parts: an in-script specification of what and how to log, and a special
function called \fct{run\_file} to run the script.  This implementation is a
direct consequence of two of the design principles mentioned in the
introduction: \emph{separation of concerns} and being \emph{agnostic}. Indeed,
there are only a few ways to implement monitoring. One is to copy the mechanism
that is used for process loggers such as \pkg{futile.logger}, and require users
to insert explicit logging expressions at multiple places within their code.
This method violates \emph{separation of concerns} as it heavily mixes data
processing code with data monitoring code.  Another way is to alter the data
processing functions so that they detect whether an object is being monitored,
at which point they make sure that monitoring code is executed. This would
violate the \emph{agnostic} principle as it implies an explicit relation
between data processing code and data monitoring code. The third way is to
intercept expressions as they are executed and insert monitoring code at
runtime. This is also what the tracing functions in base \proglang{R} do for
debugging purposes. In this sense, \pkg{lumberjack} mimics the behaviour of
base \proglang{R} tracing: it offloads the monitoring interventions to a
special `code runner' that knows what objects are monitored in which way.

\subsection{Monitoring data in interactive mode}
For logging in interactive \proglang{R} sessions, \pkg{lumberjack} defines a
special `pipe' operator, denoted \code{\%L>\%}, that can be used to chain
expressions together. When used without logging it works similar (but not
exactly the same) to the well known \pkg{magrittr} pipe operator of
\cite{bache2014magrittr}: output of the left-hand-side is fed as the first
argument to the function call on the right-hand-side.
\begin{Schunk}
\begin{Sinput}
R> spm <- read.csv("supermarkets.csv")
R> spm %L>% 
+   transform(other.rev = ifelse(is.na(other.rev), 0, other.rev )) %L>%
+   transform(ratio = turnover/total.rev) %L>%
+   head(3)
\end{Sinput}
\begin{Soutput}
     id staff turnover other.rev total.rev     ratio
1 SPM01    75       NA         0      1130        NA
2 SPM02     9     1607         0      1607 1.0000000
3 SPM03    NA     6886       -33      6919 0.9952305
\end{Soutput}
\end{Schunk}
To record what happens at each expression in the chain, a logger must be inserted
and subsequently stopped.
\begin{Sinput}
R> out <- spm %L>%
+   start_log(cellwise$new(key = "id")) %L>%
+   transform(other.rev = ifelse(is.na(other.rev), 0, other.rev)) %L>%
+   transform(ratio = turnover/total.rev) %L>%
+   stop_log()
\end{Sinput}
\begin{CodeOutput}
Dumped a log at cellwise.csv
\end{CodeOutput}
The name of the default output file is not prepended with the name of the
variable being monitored as in Section~\ref{sect:scripts}.  The reason is that
\fct{start\_log} can not in all circumstances easily determine the name of the
variable under scrutiny.  It is also of less importance, when compared to the
case presented in Section~\ref{sect:scripts}, since a chain of operations can
only process a single data object.

The log can be retrieved again by reading the log file. Below, the first and
last lines of the logging data are shown.
\begin{Schunk}
\begin{Sinput}
R> spm_log <- read.csv("cellwise.csv")
R> rbind(head(spm_log,1), tail(spm_log,1))
\end{Sinput}
\begin{Soutput}
   step                     time srcref
1     1 2020-05-08 15:24:36 CEST     NA
91    2 2020-05-08 15:24:36 CEST     NA
                                                      expression   key
1  transform(other.rev = ifelse(is.na(other.rev), 0, other.rev)) SPM01
91                         transform(ratio = turnover/total.rev) SPM60
    variable old          new
1  other.rev  NA 0.0000000000
91     ratio  NA 0.0007087172
\end{Soutput}
\end{Schunk}

Here, the logger is created with \fct{cellwise\$new} as usual. The `pipe'
operator fulfills the task of detecting whether data on the left-hand-side is
logged.  If so, it will store a copy and execute the right-hand-side with data
from the left-hand-side as input to create the output. Next, the input stored
earlier, the output, and some metadata is fed to the logger so it can measure
the difference and finally \code{\%L>\%} returns the output.  One can think of
\code{\%L>\%} is a `dressed' pipe operator that does something extra on top of
passing output of one expression as input to another (i.c., making sure that
the logging information is created).

\newpage{}
\section{Custom loggers}
\label{sect:custom}
The \pkg{lumberjack} package allows users and package authors to create custom
loggers. In order for the logger to work with \pkg{lumberjack} it must meet a
few requirements. In short, it must be a reference object with an \fct{\$add}
method for adding entries to the log, and a \fct{\$dump} method for dumping log
data. In the rest of the Section these requirements are discussed in more
detail.  It is assumed that the reader is somewhat familiar with
object-oriented programming in \proglang{R}.

Any type of reference object based on \proglang{R} environments may work but it
is recommended to use the \code{R6} system of \citet{chang2019r6} or the
\code{RefClass} system from the \pkg{methods} package \citep{rcore2019}. In the
current paper \code{R6} is used but an example using \code{RefClass} can be
found in the `extending \pkg{lumberjack}' vignette that is included with the
package.

To create a logger for lumberjack, the new \code{R6} class must have an
\fct{add} method with the following signature.
\begin{Sinput}
  $add(meta, input, output)
\end{Sinput}
The task of this method is to use the \code{input} and/or the \code{ouput} data
to create logging information and add this to the log. Optionally it can use
the information in \code{meta} to enrich the logging information.
\code{lumberjack} puts no restrictions on the data type of \code{input} and
\code{output}. It is thus possible to create loggers for any type of data. When
data is logged by a custom logger, \code{lumberjack} will make sure that the
first argument (\code{meta}) is passed a named \class{list} with two elements.
Element \code{meta\$expr} is the \proglang{R} \class{expression} that turned
\code{input} into \code{output}. Element \code{meta\$src} is the same
expression represented as a \class{character} string.  For example, the
\fct{add} method of the \code{filedump} logger (Table~\ref{tab:loggers}) just
increases an internal counter and writes \code{output} to a numbered file in a
directory.

Second, the logger must have a \fct{dump} method with the following
signature.
\begin{Sinput}
  $dump()
\end{Sinput}
It is allowed for the dump method to have extra arguments. Extra arguments
passed to \fct{stop\_log} will be passed through to the relevant \fct{\$dump}
method. For example, the \code{dump} method of the \code{cellwise} logger
accepts a \code{file} argument to specify to what file the logging information
should be exported.

In Figure~\ref{fig:trivial} the \class{trivial} logger is defined.  This logger
only registers whether data has changed at all, but it does not register which
expresssion caused the change. The final log result is therefore a simple
\code{TRUE} (object has changed) or \code{FALSE} (object has not changed). 
Althought this logger is very simple it contains all elements necessary
to define a logger.

The class definition contains one variable called \code{changed} with initial
value \code{NULL}. This is the placeholder for the logging information that
will be updated by the \fct{add} method. The \code{initialize} method is
executed when a new object of class \code{trivial} is created. At
initialization, \code{changed} is set to \code{FALSE}.
\begin{figure}
\hrule
\begin{Sinput}

  library(R6)
  trivial <- R6Class("trivial",
    public = list(
      changed = NULL
    , initialize = function(){
        self$changed <- FALSE
    }
    , add = function(meta, input, output){
      self$changed <- self$changed | !identical(input, output)
    }
    , dump = function(){
      msg <- if(self$changed) "" else "not "
      cat(sprintf("The data has %schanged\n", msg))
    }
    )
  )
\end{Sinput}
\hrule
\caption{Definition of the \class{trivial} logger using the \pkg{R6} system.}
\label{fig:trivial}
\end{figure}

Now, the \fct{add} method ignores the \code{meta} argument and sets
\code{changed} to \code{TRUE} when it already is \code{TRUE} or when
\code{input} and \code{output} are not identical. The \fct{dump} method writes
a message to screen, stating whether data has changed or not.

The code of Figure~\ref{fig:trivial} is stored in a file called
\code{trivial.R} with the supplamentary materials. Here is a demonstration of
how to use it.
\begin{Schunk}
\begin{Sinput}
R> source("trivial.R")
R> spm <- read.csv("supermarkets.csv")
R> out <- spm %L>% start_log(trivial$new()) %L>% identity() %L>% dump_log()
\end{Sinput}
\begin{Soutput}
The data has not changed
\end{Soutput}
\begin{Sinput}
R> out <- spm %L>% start_log(trivial$new()) %L>% head(10) %L>% dump_log()
\end{Sinput}
\begin{Soutput}
The data has changed
\end{Soutput}
\end{Schunk}
Here, \fct{identity} is \proglang{R}'s identity function: it just returns it's
argument unchanged. \code{head(10)} returns the first ten records data passed
to it by \code{\%L>\%}. Observe that the logger correctly notifies the user
whether the data has undergone any changes.

For some loggers it may be necessary to perform some cleanup actions when
stopping. For example, a logger may need to close a connection to a database or
remove temoprary files. For this reason one can optionally add a \fct{stop}
method. If it exists, this is called (currently with no arguments) by
\fct{stop\_log} after executing the \fct{dump} method. A typical logger object
using this construction will set up a connection object at initialisation and
close the connection when stopped.

\section{Implementation}
The techniques used to implement functionality of this package have broader use
cases then logging, and have also been documented separately in
\citet{loo2020method}. The main idea is to create a mechanism where one can
derive information from running \proglang{R} code, subject to the following
conditions. First, a user should not have to extensively edit their code in
order to create or configure the way this information is derived (a typical
counter-example is process logging where logging messages require developers to
insert logging expressions throughout their code). Second, creation or
manipulation of global variables, either in \proglang{R}'s global namespace or
in a package's namespace, e.g.\ for configuration purposes, should be avoided.
And finally, the information, derived from running \proglang{R} code, should be
transmitted through ordinary channels and not \code{stderr}. This means that
mechanisms such as (typed) error messages are to be avoided as well. The
\pkg{lumberjack} package relies on two constructions to achieve this.

The first way these objectives can be achieved is by creating a `file runner',
such as \fct{run\_file} in the \pkg{lumberjack} package. This function
parses an \proglang{R} script and runs the expressions one by one using
\proglang{R}'s \fct{parse} and \fct{eval}. This offers the possibility to
derive information from the state of the user code before and after evaluating
each expression. Since the user code is now evaluated in a custum parse-eval
loop there is also no need for using exceptions to convey logging information.
In order to capture user commands, such as those expressed by
\fct{start\_log} or \fct{dump\_log}, these functions are masked by
\fct{run\_file}. That is to say, the functions are replaced by the exact
same function as the one that the user is calling, except that they also write
some output into an \proglang{R} \code{environment} that is only accessible
from within \fct{run\_file}. Hence, the use of a global state for
configuring which variables are traced and how to trace them, is avoided. This
information is only stored within the scope of \fct{run\_file}. Furthermore,
the masking of the user-facing functions only takes place while
\fct{run\_file} is doing its work, so again no changes the global
environment are required.

The second way in which separation between logging and user code is achievied
is through the \code{\%L>\%} operator. In this case there is no masking or
custom parse and evaluation function. The idea here is that the logging object
travels with the data that is tracked. The function \fct{start\_log} returns its
argument with a new logger attached as an attribute. The \code{\%L>\%} operator
detects whether loggers are present. If so, a copy of the left-hand-side is
stored. Next, the expression on the right-hand-side is evaluated with the
left-hand-side approprately substituted. The output of this evaluation,
together with the input and some metadata are fed to the attached loggers. If
evaluation of the expression resulted in the removal of one or more loggers,
these are reattached by \code{\%L>\%}, after which the resulting data is
returned.

\section{Conclusion}
\label{sect:conclusion}
The \code{lumberjack} package allows users to monitor changes in data with
minimal coding effort, both in interactive and production (batch)
circumstances. Monitoring is specified by assigning a logger to an \proglang{R}
object, thereby separating concerns between creating data processing code and
data monitoring code. It is possible to track multiple \proglang{R} objects
simultaneously and to track an \proglang{R} object with multiple loggers.  The
tracking itself is agnostic of the code used to manipulate the objects under
scrutiny and can be used in combination with any (third party) \proglang{R}
code.  The way tracking takes place is flexible since it can be fully
customized by creating a logging object type satisfying a small set of
interface requirements.

\subsection*{Acknowledgements}
The author is indebted by Dr.\ K.\ Olav ten Bosch for carefully reading the original
manuscript.

\bibliography{jss4008}

\end{document}